\documentclass[12pt]{article}

\usepackage[final]{graphicx}
\usepackage{amsmath}
\usepackage{amssymb} % pour mathbb et mathfrak
\usepackage{oldgerm} % pour textgoth, textfrac et textswab

\title{Conditioning SLEs \\ and \\ loop erased random walks}
\date{}
\author{}

\newcommand{\Expect}[1]{{\mathbf E}\left[ #1 \right]}
\newcommand{\tExpect}[1]{\tilde{{\mathbf E}}\left[ #1 \right]}

\newcommand{\statav}[1]{\langle #1 \rangle}

\begin{document}
\maketitle

\vspace{-1.5 truecm}

\centerline{\large Michel Bauer\footnote[1]{ Institut de Physique
    Th\'eorique de Saclay, CEA-Saclay, 91191 Gif-sur-Yvette, France
    and Laboratoire de Physique Th\'eorique, Ecole Normale
    Sup\'erieure, 24 rue Lhomond, 75005 Paris, France.  {\small \tt
      <michel.bauer@cea.fr>}}, Denis Bernard\footnote[2]{Member of the
    CNRS; Laboratoire de Physique Th\'eorique, Ecole Normale
    Sup\'erieure, 24 rue Lhomond, 75005 Paris, France.  {\small \tt
      <denis.bernard@ens.fr>}}, Tom Kennedy \footnote[3]{Department of
    Mathematics, University of Arizona, Tucson, AZ 85721. {\small \tt
      <tgk@math.arizona.edu>}}} 

\vspace{.3cm}

\begin{abstract}
We discuss properties of dipolar SLE$_{\kappa}$ under conditioning.
We show that $\kappa=2$, which describes continuum limits
of loop erased random walks, is characterized as being
the only value of $\kappa$ such that dipolar SLE conditioned
to stop on an interval coincides with dipolar SLE on
that interval. We illustrate this property by computing
a new bulk passage probability for SLE$_2$. 
\end{abstract}

\newpage
\section{Introduction}

Schramm-Loewner evolution (SLE) comes in several flavors. 
Chordal SLE describes random curves between two points on 
the boundary of a simply connected domain. Radial SLE describes 
random curves between an interior point and a boundary point. 
Dipolar SLE describes random curves between a boundary point and 
a subarc of the boundary. These various types of SLE are closely 
related to each other and to the more general SLE$_{\kappa,\rho}$ 
processes \cite{Schramm-Wilson, Werner, Dubedat}

We consider dipolar SLE from a boundary point $z$ to a subarc $A$ of 
the boundary and ask what happens if we condition it to end in 
a subarc $A^\prime \subset A$, or to end at a point $w \in A$. 
We will show that the result is an SLE$_{\kappa,\rho}$ process. In 
the particular case of $\kappa=2$, when the dipolar SLE is conditioned
to end at a fixed point the distribution of the resulting curve is 
just chordal SLE.

Let us observe however that even if a discrete model is known to (or
conjectured to) converge to chordal SLE$_{\kappa}$ in the continuum
limit, there is at the moment very little understanding how to modify
it to a discrete process candidate to converge to an
SLE$_{\kappa,\rho}$ in the continuum limit. Even the definition of
``dipolar self avoiding walk'' escapes us at the moment.

Certain lattice models of random curves have special properties that
allow us to identify the associated $\kappa$ in the continuum limit
assuming the continuum limit is an SLE. The restriction property of
the self avoiding walk implies that $\kappa=8/3$ as the only possible
candidate for the scaling limit.  The locality property of percolation
identifies $\kappa=6$ as the only possible scaling limit.

The loop-erased random walk (LERW) has another property that together
with our result for $\kappa=2$ implies that if the continuum limit 
is SLE, then $\kappa$ must be $2$. 
For concreteness, we consider the model in the infinite  strip.
Consider a random walk starting at the 
origin and conditioned to exit the strip on the upper boundary. 
Loop-erasing this model gives what we will refer to as the dipolar LERW.
If we take the original random walk which was conditioned to exit 
through the upper boundary and further condition it to exit through 
a point $w$ in the upper boundary, the result is the same 
as conditioning a random walk starting at the origin to exit the 
strip through $w$. So if we condition the dipolar LERW to end at $w$, 
then the result is the same as the loop erasure of a random walk
starting at $0$ and conditioned to exit at $w$. The latter is just 
the chordal LERW. In short, conditioning the dipolar LERW to end 
at a particular point gives the chordal LERW. As we will see, SLE has 
this property only for $\kappa=2$. 

The LERW has been proved to converge to SLE$_2$ in the scaling limit
\cite{LSW,Zhan}. Since the LERW trivially has the property 
that conditioning the dipolar version to end at a fixed point 
give the chordal version, the proofs of convergence of the LERW to 
SLE$_2$ have as a corollary the result that conditioning dipolar SLE$_2$ 
to end at a fixed point gives chordal SLE$_2$. 
Nonetheless, it is desirable to have a direct proof that SLE$_2$ 
has this property.

Properties that SLE satisfies for only one value of the parameter 
$\kappa$ are useful for identifying the scaling limit of 
discrete or lattice models that satisfy the property. 
Recently, claims that SLE samples describe lines in a variety of
2d systems have appeared : these claims concern zero vorticity
lines in turbulence \cite{Bernard et al}, nodal lines of chaotic or
random wave function \cite{Bogomolny et al} and interfaces in ground
states of spin glasses \cite{Hartmann et al,Bernard et other al}. The
first two cases seem to involve $SLE_6$ (i.e. percolation).  This is
far from fully justified, even if some heuristic theoretical arguments
make this plausible. The case of spin glasses is even more intricate.
The value of $\kappa$ computed numerically is not too far (but
apparently reliably different) from $2$. Moreover there are several
definitions of the notion of interface. For some of the definitions
conformal invariance has been ruled out so that only for one
definition is the relationship with SLE still a plausible guess.

Of course numerical studies are made in finite geometries (even with
finite systems) and the problem of finite size effects has to be taken
into account. In particular dipolar SLE, which is well adapted to a
strip geometry, seems to give much more accurate numerical evaluations
of $\kappa$ and of the quadratic variation of the driving process than
chordal SLE.

SLE has been studied by researchers coming from a variety of disciplines
ranging from probability to conformal field theory.
To make this article accessible to these various groups, we
derive our results in several ways. In addition, for several results
that have already appeared in the literature we give another derivation 
from a different point of view.

\section{Avatars of dipolar and chordal SLE}

\subsection{Preliminaries}

In the half plane geometry $\mathbb{H}\equiv\{z,\; \text{Im}\; z
\in]0,+\infty[\}$, a growth process of a piece of curve starting from
a point on the lower boundary $\text{Im}\; z=0$ can be uniformized by
a map $g(z)$ having the point at infinity as a fixed point with the
normalization: $g(z) \simeq z+2c/z+o(1/z)$ for $z \rightarrow \infty$. It is
customary to call this the hydrodynamic normalization. The number $c$
is real, nonnegative, and increases during the growth. Then, with a
time parameterization $s$ of the growing trace such that $c_s\equiv
s$, the uniformizing map satisfies
\[
\frac{dg_s(z)}{ds}=\frac{2}{g_s(z)-V_s}, \qquad g_0(z)=z
\]
Chordal SLE from $0$ to $\infty$ is the stochastic growth
process obtained when $V_s=\sqrt{\kappa}B_s$ where $B_s$ is a standard
Brownian motion.  

\vspace{.3cm}

In the strip geometry $\mathbb{S}\equiv\{z,\; \text{Im}\; z
\in]0,\pi[\}$ a growth process of a piece of curve starting from a
point on the lower boundary $\text{Im}\; z=0$ can be uniformized by a
map $g(z)$ fixing the two points at infinity. This leaves the
possibility to translate $g$ by a real constant. But $g(z) \simeq z
+c_{\pm}+o(1)$ for $z \rightarrow \pm \infty$, and $g$ can be fully
normalized by requiring that $c_++c_-=0$. We propose to call this the
strip symmetric normalization. Then, with a time parameterization $s$
such that $c_+=s$ of the growing trace, the uniformizing map satisfies

\[
\frac{dg_s(z)}{ds}=\frac{1}{\tanh (g_s(z)-V_s)/2}, \qquad g_0(z)=z
\]
Throughout the paper $\tanh x/2$ stands for $\tanh (x/2)$. Dipolar
SLE$_{\kappa}$ from $0$ to the upper boundary $\text{Im}\; z =\pi$ (as
defined in \cite{Bauer Bernard Houdayer}) is the stochastic growth
process obtained when $V_s=\sqrt{\kappa}B_s$ where $B_s$ is a standard
Brownian motion.

\subsection{Change of domain} \label{subsec:change of domain}

Chordal SLE is particularly simple in the half plane geometry with the
hydrodynamic normalization and dipolar SLE is particularly simple in
the strip geometry with the symmetric normalization. But both are
conformally invariant processes, and as such could be described in any
domain (with two marked points for the chordal case, or with one
marked boundary interval and one marked point in its boundary
complement in the dipolar case) and moreover with any choice of
normalization of the Loewner map.

We illustrate this issue with four examples. The first one is well
known. We treat the second one in detail, using a commutative diagram
technique introduced in the study of locality and restriction for SLE,
see e.g. \cite{Lawler et other al}, and sketch the treatment of the
fourth. The same method can be used in all four cases, leading to
straightforward but slightly painful computations. We shall see later
how ideas from statistical mechanics and conformal field theory
followed by a simple application of Girsanov's theorem allow to
simplify the computations.

\vspace{.2cm}

\textbf{Chordal SLE from $0$ to $a$ in the upper half plane} with the
hydrodynamic normalization is described by the system of stochastic
differential equations:
\[\frac{dg_s(z)}{ds}=\frac{2}{g_s(z)-V_s}, \quad
\frac{dA_s}{ds}=\frac{2}{A_s-V_s},\] 
\[ dV_s=\sqrt{\kappa}dW_s+(\kappa-6)\frac{ds}{V_s-A_s},\]
with initial conditions $g_0(z)=z$, $V_0=O$ and $A_0=a$.

\vspace{.2cm}

\textbf{Chordal SLE from $0$ to $i\pi+a$ in the strip} with the
symmetric normalization is described by the system of stochastic
differential equations:
\[\frac{dg_s(z)}{ds}=\frac{1}{\tanh (g_s(z)-V_s)/2} , \quad
\frac{dA_s}{ds}=\tanh (A_s-V_s)/2,\]
\[ dV_s=\sqrt{\kappa}dW_s+(\kappa/2-3)ds\tanh (V_s-A_s)/2  \] 
with initial conditions $g_0(z)=z$, $V_0=O$ and $A_0=a$.

\vspace{.2cm}

\textbf{Dipolar SLE from $0$ to $[a,b]$ in the upper half plane} with the
hydrodynamic normalization is described by the system of stochastic
differential equations:
\[\frac{dg_s(z)}{ds}=\frac{2}{g_s(z)-V_s} , \quad
\frac{dA_s}{ds}=\frac{2}{A_s-V_s} , \quad \frac{dB_s}{ds}=\frac{2}{B_s-V_s},\]
\[ dV_s=\sqrt{\kappa}dW_s+(\kappa/2-3) \left(\frac{ds}{V_s-A_s} +
\frac{ds}{V_s-B_s}\right),\] 
with initial conditions $g_0(z)=z$, $V_0=O$, $A_0=a$ and $B_0=b$.

\vspace{.2cm}

\textbf{Dipolar SLE from $0$ to $[i\pi+a,i\pi+b]$  in the strip} with the
symmetric normalization is described by the system of stochastic
differential equations:
\begin{eqnarray*} \frac{dg_s(z)}{ds} & = & \frac{1}{\tanh
    (g_s(z)-V_s)/2}, \\
\frac{dA_s}{ds}& = & \tanh (A_s-V_s)/2, \quad \frac{dB_s}{ds}\; = \;\tanh
(B_s-V_s)/2,\\
dV_s & = & \sqrt{\kappa}dW_s+(\kappa/2-3)ds\frac{\tanh (V_s-A_s)/2 + \tanh
(V_s-B_s)/2}{2}\end{eqnarray*}
with initial conditions $g_0(z)=z$, $V_0=O$, $A_0=a$ and
$B_0=b$.

\vspace{.4cm}

In all the above systems, $W_s$ is a normalized Brownian
motion. Before we explain the second case, let us make a few remarks.\\
-- The drifts all vanish at $\kappa=6$, which is a
manifestation of the locality of percolation.\\
-- For $\kappa \leq 4$ the above equations give a complete description.
For $\kappa >4$, the first two equations give only a partial
description : the first is a description of chordal SLE from $0$ to
$a$ in the upper half plane only up to the first time the process
separates $a$ from $\infty$ by hitting the real line in the interval
$[a,\infty]$ that does not contain $0$; the second is a description of
chordal SLE from $0$ to $i\pi+a$ in the strip only up to the first
time the process hits the line $\text{Im}\; z =\pi$. In both cases,
the trouble comes solely from the choice of normalization.

\vspace{.5cm}

With this proviso in mind, we go on to write down chordal SLE in
the strip with the symmetric normalization. The starting point will be
$0$ and the end point will be $i\pi+a$ where $a$ is a real number.  An
invaluable tool to make computations straightforward is a commutative
diagram, just as in the study of locality and restriction for SLE
\cite{Lawler et other al}.

\begin{figure}
\begin{center}
\includegraphics[height=7cm]{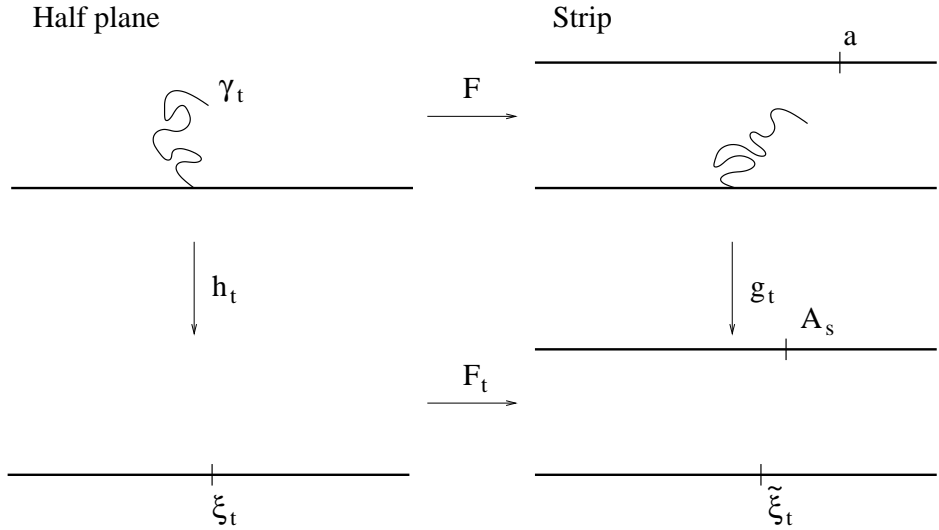}
\end{center}
\caption{The basic commutative diagram}
\end{figure}

Uniformize chordal SLE in the upper-half plane from $0$ to $\infty$
with trace $\gamma_{[0,t]}$ up to time $t$ by
$h_t:\mathbb{H}\backslash \gamma_{[0,t]} \rightarrow \mathbb{H}$ in the
hydrodynamical normalization:
\[ \frac{dh_t(w)}{dt}=\frac{2}{h_t(w)-\xi_t}, \quad
\xi_t=\sqrt{\kappa}B_t.\] Map the upper half plane to the strip by any
conformal map $F:\mathbb{H}\rightarrow \mathbb{S}$ such that $F(0)=0$
and $F(\infty)=i\pi+a$ to get SLE in $\mathbb{S}$ from $0$ to
$i\pi+a$. The formula for $F$ is
\[ \tanh (F(w)-a)/2 +\tanh a/2 =lw.\]
where $l \in \mathbb{R}^{+*}$ is arbitrary (we have given only two
conditions to normalize $F$).
Let $g_t$ be the uniformizing map $g_t:\mathbb{S}\backslash
F(\gamma_{[0,t]}) \rightarrow \mathbb{S}$ with the symmetric normalization
as above except that the time parameterization is not at our disposal 
\[ \frac{dg_t(z)}{dt}=\frac{a_t}{\tanh (g_t(z)-\tilde{\xi}_t)/2}.\]
Finally let $F_t:\mathbb{H}\rightarrow
\mathbb{S}$ be the map ``closing the square'',
\[ g_t \circ F = F_t \circ h_t.\]

\vspace{.1cm}

Take the time derivative
of $g_t \circ F = F_t \circ h_t$ and afterwards substitute $w$ for
$h_t$ to get
\[\frac{a_t}{\tanh (F_t(w)-\tilde{\xi}_t)/2}=
\frac{dF_t(w)}{dt}+F'_t(w)\frac{2}{w-\xi_t}.\] 
Now $F_t(w)$ is non singular at $w=\xi_t$ so that the pole on the
right hand-side is cancelled by a pole on the left hand-side:
\[\frac{a_t}{\tanh (F_t(w)-\tilde{\xi}_t)/2}=F'_t(\xi_t)\frac{2}{w-\xi_t} 
+ O(1) \text{ when } w\rightarrow \xi_t,\]
leading to $\tilde{\xi}_t=F_t(\xi_t)$ and
$a_t=F'_t(\xi_t)^2$. Continuing the expansion one step further yields
\[-a_tF''_t(\xi_t)/F'_t(\xi_t)^2=\frac{dF_t}{dt}(\xi_t)+2F''_t(\xi_t)
\text{ i.e. } \frac{dF_t}{dt}(\xi_t)=-3F''_t(\xi_t).\]
Now use Ito's formula (in a slightly extended context as usual in this
computation) to get
\[d\tilde{\xi}_t=d(F_t(\xi_t))=dF_t(\xi_t)+F'_t(\xi_t)d\xi_t+
\frac{\kappa}{2} F''_t(\xi_t)dt,\]
i.e. $d\tilde{\xi}_t=F'_t(\xi_t)d\xi_t+(\kappa /2-3)F''_t(\xi_t)dt$. 
Define $s(t)\equiv\int_0^t F'_u(\xi_u)^2du$ and
$\sqrt{\kappa}W_{s(t)}\equiv \int_0^t F'_u(\xi_u)d\xi_u$ so that
$W_{s}$ is a standard Brownian motion with time parameter $s$. Set
$V_{s(t)} \equiv F_t(\xi_t)$ so
\[dV_s=\sqrt{\kappa}dW_s+(\kappa
/2-3)\frac{F''_{t(s)}(\xi_{t(s)})}{F'_{t(s)}(\xi_{t(s)})^2}ds.\]

This equation is in fact valid for all four cases, but of course $F_t$
closes a different commutative diagram in each case. Computations are
simplified a little bit when one realizes that if $F(w)=z$ has inverse
$\Phi(z)=w$ then 
\[ \frac{F''(w)}{F'(w)^2}=-(\log \Phi'(z))'.\] In this identity, the $'$
indicates derivative with respect to $w$ on the left hand side but with
respect to $z$ on the right hand side.  So if we write $\Phi_s(z)=w$
for the inverse of $F_t(w)=z$ (where $s$ and $t$ are related by the
time change) we get the important intermediate formula :
\begin{equation} \label{eq:drift1} 
dV_s=\sqrt{\kappa}dW_s+(3-\kappa /2)(\log \Phi_s'(z))'_{|z=V_s} ds.
\end{equation}

\vspace{.1cm}

Let $A_s \equiv g_{t(s)}(i\pi+a)-i\pi$ be the trajectory of $a$
under the Loewner evolution. From $g_t \circ F = F_t \circ h_t$,
$h_t(\infty)=\infty$ and $F(\infty)=i\pi+a$ we learn that
$g_t(i\pi+a)=F_t(\infty)=i\pi+A_s$. Combined with
$F_t(\xi_t)=V_{s(t)}$ we find
\[\tanh (F_{t(s)}(w)-A_s)/2 +\tanh (A_s-V_s)/2 =l_s(w-\xi_{t(s)}),\]
or
\[\tanh (z-A_s)/2 +\tanh (A_s-V_s)/2 =l_s(\Phi_s(z)-\xi_{t(s)}),\]
where $l_s$ is an unspecified but irrelevant parameter. 

The computation of the drift is now easy (the reader should compare
with a direct use of the formula involving $F_t$) :
\[((\log \Phi_s'(z))'_{|z=V_s}=-\tanh (V_s-A_s)/2.\]

\vspace{.1cm}

This finishes to establish the formula announced above for chordal SLE in
the strip from $0$ to $i\pi+a$ in the symmetric normalization. 

\vspace{.3cm}

We end this section with some remarks on how to write down dipolar SLE
from $0$ to $[i\pi+a,i\pi+b]$ in the strip with the symmetric
normalization. The steps above could be reproduced, leading to
equation (\ref{eq:drift1}), but with a function $\Phi_s$ mapping the
strip to the strip, $V_s$ to $\xi_{t(s)}$, $i\pi+A_s$ to $-\infty$ and
$i\pi+B_s$ to $+\infty$. This fixes $\Phi_s$ uniquely. A convenient
representation is 
\[ \coth (\Phi_s(z)-\xi_{t(s)})/2=\nu + \mu \coth (z-V_s)/2\]
with  
$-1=\nu + \mu \tanh (A_s-V_s)/2$, $1= \nu + \mu \tanh (B_s-V_s)/2.$

To compute the drift we have to be able to take two derivatives with
respect to $z$ and then put $z=V_s$ so it is enough to use the second
order expansion:
\[\frac{\Phi_s(z)-\xi_{t(s)}}{2}=\frac{z-V_s}{2\mu}-
\nu\left(\frac{z-V_s}{2\mu}\right)^2+\cdots\] 
for $z \rightarrow V_s$. We infer that
\[(\log \Phi_s'(z))'_{|z=V_s}=-\nu/\mu = \frac{\tanh (V_s-B_s)/2 +
  \tanh (V_s-B_s)/2 }{2},\] 
leading to the announced formula.

\section{Conditioning}

We aim at a description of dipolar SLE in a domain (call it $D$) from
a boundary point (call it $0$) to a boundary interval (call it $I$)
conditioned to hit either a subinterval $J$ of $I$ or, via a limiting
procedure, a point in $I$.

\subsection{Na\"{\i}ve considerations}
\label{subsec:naive}

Girsanov's theorem is the general tool to tackle these problems, but
in this section we want to illustrate what goes on using only
elementary manipulations, with the hope that the explicit argument
shows in the clearest way what is involved, compensating at least
partly the lack of elegance of the approach.

\vspace{.2cm}

We could in principle make computations in any domain, with any
normalization of the uniformizing maps. For illustration, we choose
$D$ to be the strip $\mathbb{S}$, $I$ to be upper boundary
$\text{Im}\; z =\pi$, and $J$ to be the interval $]i\pi+a,i\pi+b[$
where $a < b$ are real numbers. Before conditioning, the process is
described by the equations
\[
\frac{dg_s(z)}{ds}=\frac{1}{\tanh (g_s(z)-V_s)/2}, \qquad g_0(z)=z
\]
where $V_s=\sqrt{\kappa}W_s$ and $W_s$ is a standard
Brownian motion. The inverse image $g_s^{-1}(\mathbb{S})$ is
$\mathbb{S}\backslash \gamma(]0,s])$. In particular, the inverse
image of $g_s^{-1}(V_s)$ --defined as $\lim_{\varepsilon \mapsto
  0^+}g_s^{-1}(V_s+i\varepsilon)$-- is $\gamma(s)$. 

\vspace{.2cm}

\textbf{The martingale} We condition on the event that the curve ends
in an interval $J$. We will denote this event by $J$ as well, no
confusion should arise. The first thing we have to know is the
probability of $J$. This is a routine computation if one uses
martingales, but this is also the key to make contact with Girsanov's
theorem later. Write $p(J)$ for the hitting probability of dipolar SLE
and observe that $p(J)=f(a)-f(b)$ where $f(c)$ is the probability to
hit on the right of $c$. So it is enough to deal with the case when
$J=]i\pi+a,i\pi+\infty[$. Observe that $A_s\equiv g_s(i\pi+a)-i\pi$
satisfies
\[
\frac{dA_s}{ds}=\tanh (A_s-V_s)/2 , \quad A_0=a. \] 

Our claim is the following: for $t \geq 0$,
$\Expect{1_J|\mathcal{F}_t}=f(A_t-V_t)$. Recall that an
$\mathcal{F}_t$-measurable event is an event whose realization can be
decided from the knowledge of the process $W_{\bullet}$ for times up
to $t$.  So the left hand side is the probability to hit in $J$
knowing the process $W_{\bullet}$ up to time $t$. Intuitively, this
knowledge is equivalent to the knowledge of $g_t$. Dipolar SLE is
conformally invariant, and in particular, for $s \geq t$, the
statistics of $g_t(\gamma(]t,s])-V_t$, which is supposed to hit on the
right of $A_t-V_t$, is the same as the statistics of a
$\gamma(]0,s-t])$. The hitting probability of this process is
$f(A_t-V_t)$ as announced.

The equation $\Expect{1_J|\mathcal{F}_t}=f(A_t-V_t)$ means that
$f(A_t-V_t)$ is a --so called ``closed''-- martingale. In particular
the Ito derivative of the stochastic process $f(A_t-V_t)$ has no drift. A
routine use of Ito's formula yields
\[\frac{\kappa}{2}f''(a)+ f'(a) \tanh a/2 =0,\]
whose only solution with an acceptable probabilistic interpretation is
proportional to $\int_a^{+\infty}dx \cosh ^{-4/\kappa} x/2$, leading to 
\[p(J)=\frac{\int_a^b \, dx \cosh ^{-4/\kappa} (x/2)}
{\int_{-\infty}^{+\infty}\, dx \cosh ^{-4/\kappa} (x/2)} \]
for the hitting probability of the interval $J=]i\pi+a,i\pi+b[$, a
formula established in \cite{Bauer Bernard Houdayer}. 

\vspace{.2cm}

\textbf{Markov property} Let $p$ denote the probability distribution for
dipolar SLE (more precisely for the driving process $V_s$, which is
proportional to a Brownian motion), and let $\tilde{p}$ denote the
probability distribution for dipolar SLE conditioned to hit the upper
boundary in $J$ (again more precisely for the driving process
$V_s$, which after conditioning is not simply a Brownian motion
anymore). Define $J_s \equiv g_s(J)=]i\pi+A_s,i\pi+B_s[$.  

Observe that the knowledge of $V_t,A_t,B_t$ for all $t$'s is a
(redundant) description of the growth process. It is redundant because
the knowledge of $V_t$ for all $t$'s would suffice. For $A < B$ and $t
\geq 0$, let $E_{t,V,A,B}$ denote the set of 
driving functions $V_{\bullet}$ such that $V_t\in [V,V+dV]$ $A_t\in
[A,A+dA]$ and $B_t \in [B,B+dB]$. This is an
$\mathcal{F}_t$-measurable set. Let $E$ be any
$\mathcal{F}_t$-measurable subset of $E_{t,V,A,B}$, and for $t \leq s$
compute $\tilde{p}(E \text{ and } E_{s,V',A',B'})$. By definition,
this is
\[\frac{p(E \text{ and } E_{s,V',A',B'} \text{ and } \gamma \text{ hits in }
]a,b[)}{p(\gamma \text{ hits in }]a,b[)}.\]
But under $p$, the process $(V_{\bullet},A_{\bullet},B_{\bullet})$ is
Markovian, so that 
\[p(E \text{ and } E_{s,V',A',B'} \text{ and }
\gamma \text{ hits in } ]a,b[)=p(E)p(E')p(J')\] where $E'$ is the
event that starting from $V,A,B$ at time $t$,
$V_s\in[V',V'+dV'],A_s\in[A',A'+dA'],B_s\in[B',B'+dB']$ and $J'$ is
the event that dipolar SLE started at $V_s$ on the lower boundary hits
the upper boundary in the interval $J'=]A',B'[$, or by translation
invariance, that  dipolar SLE started at $0$  on the lower boundary hits
the upper boundary in the interval $J'=]A'-V',B'-V'[$. We have already
computed $p(J')$. Hence
\[\tilde{p}(E \text{ and } E_{s,V',A',B'})=p(E)p(E')\frac{p(\gamma
  \text{ hits in }]A'-V',B'-V'[)}{p(\gamma \text{ hits in }]a,b[)}.\]
A simpler argument yields
\[\tilde{p}(E)=p(E)\frac{p(\gamma\text{ hits in
  }]A-V,B-V[)}{p(\gamma \text{ hits in }]a,b[)},\]  
and by comparison,
\[ \tilde{p}(E \text{ and } E_{s,V',A',B'})=
\tilde{p}(E)p(E')\frac{p(\gamma \text{ hits in
  }]A'-V',B'-V'[)}{p(\gamma \text{ hits in }]A-V,B-V[)}\] 
for any $\mathcal{F}_t$-measurable event $E \subset E_{t,V,A,B}$.

This proves that under $\tilde{p}$, the process $V_t,A_t,B_t$ is
Markovian and, from the definition of $E'$, that the transition
probability density under $\tilde{p}$ to go from $(V,A,B)$ at time $t$
to $(V',A',B')$ at time $s$ is simply the transition probability
density under $p$ to go from $(V,A,B)$ at time $t$ to $(V',A',B')$ at
time $s$ weighted by the known ratio $\frac{p(\gamma \text{ hits in
  }]A'-V',B'-V'[)}{p(\gamma \text{ hits in }]A-V,B-V[)}$. Under $p$,
$V_t$ alone is a Markov process, so there is some price to pay to save
the Markov property.

\vspace{.2cm}

\textbf{Computation of the drift} Note that the transition probability
density under $p$ to go from $(V,A,B)$ at time $t$ to $(V',A',B')$ at
time $s$ is translation invariant and time homogeneous, so that we can
assume without loss of generality that $V=0$ and $t=0$. Now we assume
that $s=\varepsilon$ and $V'=\Delta$ are small and we compute the
probability distribution of $\Delta$. Under $p$, the density of
$\Delta$ is $\frac{1}{\sqrt{2\pi\kappa\varepsilon}}
e^{-\Delta^2/(2\kappa\varepsilon)}$, and the argument in the
exponential is of order $1$ only when $\Delta \sim
\sqrt{\varepsilon}$. The conditioning ratio $\frac{p(\gamma \text{
    hits in }]A'-V',B'-V'[)}{p(\gamma \text{ hits in }]A-V,B-V[)}$ can
be expanded using this scaling, and only terms up to order $\Delta$
need to be kept. As $A'-A$ and $B'-B$ are of order $\varepsilon$ we
find
\[
\frac{p(\gamma \text{ hits in }]A'-V',B'-V'[)}{p(\gamma \text{ hits in
  }]A-V,B-V[)}=1-\Delta F(A,B)+0(\varepsilon,\Delta^2),
\]
where 
\[ F(A,B)=\frac{\cosh ^{-4/\kappa} B/2 - \cosh ^{-4/\kappa} A/2}
{\int_A^B \, dx \cosh ^{-4/\kappa} x/2} \]

So the transition probability density under $\tilde{p}$ of
$V_{\varepsilon}$ starting from $V_0=0$ is 
\[\frac{1}{\sqrt{2\pi\kappa\varepsilon}}
e^{-(\Delta+\kappa\varepsilon F(A,B))^2/(2\kappa\varepsilon)}
(1+0(\varepsilon,\Delta^2)).\]
The meaning of this is that for infinitesimal $\varepsilon$, 
$V_{\varepsilon}$ is a Gaussian random variable of mean
$-\kappa\varepsilon F(A,B)$ and standard deviation
$\sqrt{\kappa\varepsilon}$. Using translation invariance and time
homogeneity, this is exactly the meaning of the
stochastic differential equation
\[dV_s=\sqrt{\kappa}dW_s-\kappa F(A_s-V_s,B_s-V_s)ds.\] 

When $B\rightarrow A$, $F(A,B)$ has a finite limit,
$-\frac{2}{\kappa}\tanh A/2$.

\vspace{.2cm}

We summarize the central results of this section:

\textbf{Dipolar SLE from $0$ to the upper boundary in the strip,
  conditioned to hit in $[i\pi+a,i\pi+b]$} with the symmetric
normalization is described by the system of stochastic differential
equations:
\begin{eqnarray*}
\frac{dg_s(z)}{ds}& = &\frac{1}{\tanh (g_s(z)-V_s)/2}, \\
\frac{dA_s}{ds} & = &\tanh (A_s-V_s)/2 , \quad \frac{dB_s}{ds}\; =
\; \tanh
(B_s-V_s)/2,\\
dV_s & = & \sqrt{\kappa}dW_s-\kappa\left(\frac{\cosh ^{-4/\kappa} (B_s-V_s)/2 -
    \cosh ^{-4/\kappa} (A_s-V_s)/2} {\int_{A_s-V_s}^{B_s-V_s} \, dx
    \cosh ^{-4/\kappa} 
    (x/2)}\right) \end{eqnarray*}
with initial conditions $g_0(z)=z$, $V_0=O$, $A_0=a$ and $B_0=b$.

\vspace{.2cm}

\textbf{Dipolar SLE from $0$ to the upper boundary in the strip
  conditioned to hit at $i\pi+a$} with the symmetric normalization is
described by the system of stochastic differential equations:
\[\frac{dg_s(z)}{ds}=\frac{1}{\tanh (g_s(z)-V_s)/2} , \quad
\frac{dA_s}{ds}=\tanh (A_s-V_s)/2,\]
\[ dV_s=\sqrt{\kappa}dW_s-2\tanh (V_s-A_s)/2  \] 
with initial conditions $g_0(z)=z$, $V_0=O$ and $A_0=a$.

\vspace{.2cm}
 
As a corollary, dipolar SLE conditioned to hit in a subinterval is
dipolar SLE if and only if $\kappa=2$, and dipolar SLE conditioned to
hit at a given point is chordal SLE if and only if $\kappa=2$.

This is obtained by direct comparison with the formul\ae\ in
subsection \ref{subsec:change of domain}.

\subsection{An elementary application at $\kappa=2$}

The above result is a nice characterization of the value $\kappa=2$.
We show in this section that it allows us to compute in an elementary way
the probability that a dipolar SLE$_2$ trace passes to the right of a
given bulk point. The same strategy can be used to compute other
physical observables of interest, see \cite{Hagendorf}.

\vspace{.2cm}

The probability that a dipolar SLE$_{\kappa}$ trace (or hull) ends to
the right of a given boundary point of the hitting interval is
obtained by routine martingale techniques, see e.g. \cite{Bauer
  Bernard Houdayer}.

For $\kappa < 8$, the probability that the chordal SLE$_{\kappa}$
trace passes to the right of a given bulk point $z$ (for $4 < \kappa <
8$ we means that when $z$ is swallowed it goes to the negative real
axis) is again obtained via a routine martingale technique see
\cite{Schramm}.

For $\kappa \geq 4$, the probability that a dipolar SLE$_{\kappa}$
hull contains the bulk point $z$ and the probability that it passes to
the right of $z$ can be computed (somewhat miraculously) again by
routine martingale techniques by making the ansatz that they are
harmonic functions of $z$ and then checking that appropriate boundary
conditions can be imposed, see again \cite{Bauer Bernard Houdayer}.
But the harmonic ansatz fails for $\kappa < 4$.

\vspace{.2cm}

For $\kappa=2$, we obtain the probability that a dipolar SLE$_2$ trace
passes to the right of a given bulk point by the following steps.

We start from Schramm's result at $\kappa=2$ for the chordal case in
the upper half plane: if $\vartheta\in [0,\pi]$ is the argument of
$z\in \mathbb{H}$, the probability that chordal SLE$_2$ passes to the
right of $z$ is
\[\frac{\vartheta}{\pi}-\frac{\sin 2\vartheta}{2\pi}.\]

We consider a conformal map from $\mathbb{H}$ to the strip
$\mathbb{S}$ sending $0$ to $0$ and $\infty$ to the point $a+i\pi$ on
the hitting interval, to get the probability that chordal SLE$_2$ in
the strip from $0$ to $a+i\pi$ passes to the right of $w\in
\mathbb{S}$. Cumbersome but elementary algebra shows that this
probability is obtained by substitution of $\vartheta (a,w)$ for
$\vartheta$ in the above formula, where
\[ \vartheta (a,w) \equiv \arctan \frac{\sin v}{\sinh u-(\cosh u -
  \cos v) \tanh a/2},\] the function $\arctan$ takes values in
$[0,\pi]$ and $w=u+iv$ is the decomposition into real and imaginary
parts.

The trick now is that because  dipolar SLE$_2$ conditioned to hit
at $a$ is nothing but chordal SLE$_2$ aiming at $a$, the probability
$p_2(w)$ that dipolar SLE$_2$ passes to the right of $w$ can be
represented as the integral over $a$ of the dipolar hitting point
density at $a$ times the probability that chordal SLE$_2$ aiming at
$a$ passes to the right of $w$. Explicitly:
\[p_2(w)=\int_{-\infty}^{+\infty}\frac{da}{4\cosh^2 a/2}
\left(\frac{\vartheta(a,w)}{\pi}-\frac{\sin
    2\vartheta(a,w)}{2\pi}\right). 
\]
Choosing $\tanh a/2$ as the new integration variable leads to an
elementary integral because the differential of $x \arctan 1/x$ is
exactly $dx (\arctan 1/x-\frac{1}{2} \sin 2 \arctan 1/x)$. Finally:
\[p_2(w)=\frac{(e^u -\cos v) \arctan \frac{\sin v}{e^u -\cos v} -
  (\cos v -e^{-u}) \arctan \frac{\sin v}{\cos v -e^{-u}}}{2\pi (\cosh
  u - \cos v) }.\] 
with $w=u+iv$ the decomposition into real and imaginary parts.

\vspace{.2cm}

For other values of $\kappa$, we would have to work with dipolar SLE
conditioned to hit a point, and compute probability that the trace
passes to the right of $w$ for this process. This is likely to be even
more complicated than $p_{\kappa}(w)$, the probability we are aiming
at: it involves four boundary points (the starting point, the hitting
interval and the conditioned endpoint) whereas $p_{\kappa}(w)$
involves only three. The miracle of $\kappa=2$ is that after
conditioning the hitting interval plays no role anymore. Another
natural explanation for this comes from conformal field theory.

% \vspace{.2cm}

% (Do we want to elaborate a bit on that ?) There is also
% a relationship between chordal and dipolar loop erased random walks in
% the discrete setting. The following discussion is inspired by
% \cite{Zhan}. Consider the graph associated to the square lattice in
% the plane and a closed curve made of a succession of edges and
% splitting the plane in two connected pieces zzzzzzzzzzzzzzzzzzzzzzz

\subsection{Another computation using martingales}

\label{sec:sw}

Reference \cite{Schramm-Wilson} contains several computations
analogous to the above ones (some are even assigned as
exercises). In this section give another derivation of the result from
subsection \ref{subsec:naive}.

For definiteness, we consider dipolar SLE in the half plane from $0$
to $[-1,1]^c$.  (By $[-1,1]^c$ we mean the complement of $[-1,1]$ with
respect to the real axis.) The hitting density is proportional to 
\begin{eqnarray*}
  p(x)=x^{-4/\kappa} (x^2-1)^{-1 + 2/\kappa} 
  \end{eqnarray*}

We use the following martingale. For real forcing points
$x_1,\cdots,x_n$ with weights $\rho_1,\cdots,\rho_n$, it is
\begin{eqnarray*} M_t(x_1,\cdots,x_n;\rho_1,\cdots,\rho_n) =
  \prod_{j=1}^n \, L(x_j,\rho_j) \prod_{1 \le i < j \le n}
  Q(x_i,x_j;\rho_i,\rho_j)  \end{eqnarray*} 
where
\begin{eqnarray*} L(x_j,\rho_j)= |g^\prime_t(x_j)|^{\alpha_j}
  |X^{x_j}_t|^{\rho_j/\kappa} \end{eqnarray*} 
\begin{eqnarray*}
  Q(x_i,x_j;\rho_i,\rho_j) = |X^{x_i}_t - X^{x_j}_t|^{\rho_i \rho_j
    /(2 \kappa)} \end{eqnarray*} 
\begin{eqnarray*} \alpha_j=
  (8-2\kappa+2 \rho_j) \rho_j/(8 \kappa)  
\end{eqnarray*} 
Here $X^x_t$ denotes $g_t(x) - W_t$.  Let $\hat{M}_t$ be this
martingale normalized so its expectation is $1$.  Note that $M_0$ is a
constant, so $\hat{M}_t = M_t/M_0$.  If we take chordal SLE in the
half plane from $0$ to $\infty$ and weight it by $\hat{M}_t$, then we
get the SLE$_{\kappa,\rho}$ process with forcing points at the $x_i$
and weights $\rho_i$. This martingale appears in
\cite{Schramm-Wilson,Kytola}\footnote{Schramm and Wilson comment that the
  martingale was also discoverd independently by M. Biskup.}. 

Dipolar SLE in the upper half plane from $0$ to $[-1,1]^c$ is the same
as SLE$_{\kappa,\rho}$ with two force points at $-1$ and $+1$ with both
weights equal to $(\kappa-6)/2$.  This is the same as taking chordal
SLE in the upper half plane from $0$ to $\infty$ and weighting it by
the martingale $M_t(-1,1;(\kappa-6)/2,(\kappa-6)/2)$.  We want to show
that if we condition this to end at some $a \in [-1,1]$, then we get
an SLE$_{\kappa,\rho}$ process with three force points.  One is at $a$
with weight $-4$ and the other two are at $\pm 1$, both with weight
$\frac{\kappa-2}{2}$.  We will show this by showing that
\begin{eqnarray*}
  \int_{[-1,1]^c} \, 
  \hat{M}_t(a,-1,1;-4,\frac{\kappa-2}{2},\frac{\kappa-2}{2}) \,
  p(a) \, da 
  = \hat{M}_t(-1,1;\frac{\kappa-6}{2},\frac{\kappa-6}{2}) 
\end{eqnarray*}
This equation is trivially true at $t=0$.  We note that $
M_0(a,-1,1;-4,\frac{\kappa-2}{2},\frac{\kappa-2}{2}) = c p(a)$.  for
some constant $c$ which only depends on $\kappa$, and
$M_0(-1,1;\frac{\kappa-6}{2},\frac{\kappa-6}{2})$ only depends on
$\kappa$. So with some rearranging, the above is the same as
\begin{eqnarray}
  && \int_{[-1,1]^c} \, L(a,-4) \, Q(a,-1;-4,\frac{\kappa-2}{2}) \, 
  Q(a,1;-4,\frac{\kappa-2}{2}) \, \, da 
  \nonumber \\
  &=& C \, \frac{L(-1,\frac{\kappa-6}{2}) \, L(1,\frac{\kappa-6}{2}) \,
    Q(-1,1;\frac{\kappa-6}{2},\frac{\kappa-6}{2})
    }{
    L(-1,\frac{\kappa-2}{2}) \, L(1,\frac{\kappa-2}{2}) \,
    Q(-1,1;\frac{\kappa-2}{2},\frac{\kappa-2}{2}) }
\label{inteq}
\end{eqnarray}

For $\rho=(\kappa-2)/2$ and $\rho=(\kappa-6)/2$ we get the same value
of $\alpha$. Thus
\begin{eqnarray*} 
  \frac{L(p,\frac{\kappa-6}{2})}{L(p,\frac{\kappa-2}{2})}
  =  |X^p_t|^{-2/\kappa} 
\end{eqnarray*}
for $p= \pm 1$.  We find that
\begin{eqnarray*} \frac{
    Q(-1,1;\frac{\kappa-6}{2},\frac{\kappa-6}{2})}{
    Q(-1,1;\frac{\kappa-2}{2},\frac{\kappa-2}{2})} = |X_t^{-1} -
  X_t^1|^{-1+4/\kappa}
\end{eqnarray*}
We also have
\begin{eqnarray*}
  L(a,-4)  = |g^\prime_t(a)| \, |X^a_t|^{-4/\kappa}
\end{eqnarray*}
and for $p= \pm 1$,
\begin{eqnarray*}
  Q(a,p;-4,\frac{\kappa-2}{2}) = |X^a_t-X^p_t|^{(2-\kappa)/\kappa}
\end{eqnarray*}

Thus eq.(\ref{inteq}) becomes
\begin{eqnarray*}
  && \int_{[-1,1]^c} \, |g^\prime_t(a)| \, |X^a_t|^{-4/\kappa} \,
  |X^a_t-X^{-1}_t|^{(2-\kappa)/\kappa} \,
  |X^a_t-X^1_t|^{(2-\kappa)/\kappa} \, da  
  \\
  &=& c \, |X^{-1}_t|^{-2/\kappa} \, |X^1_t|^{-2/\kappa} \,
  |X_t^{-1} - X_t^1|^{-1+4/\kappa}
\end{eqnarray*}
Note that $|g^\prime_t(a)|=\frac{dX^a_t}{da}$.  $X^a_t$ is a
differentiable function of $a$ and we can do a change of variables to
change the above to
\begin{eqnarray*}
  \int_{[P,Q]^c} \, 
  X^{-4/\kappa} \,
  \left[ (X-P) (Q-X)\right]^{(2-\kappa)/\kappa}
  \, d X
  = c \, P^{-2/\kappa} \, Q^{-2/\kappa} \, (Q-P)^{-1+4/\kappa}
\end{eqnarray*}
where $X$ is shorthand for $X_t^a$, $P$ is for $X^{-1}_t$ and $Q$ for
$X^1_t$. This identity follows from the substitution
\begin{eqnarray*}
  X=PQ \frac{z+1}{Pz+Q}
\end{eqnarray*}
followed by some nice cancellations.

\section{ Girsanov's theorem}

Girsanov's theorem gives a unified view of the computations of the 
two preceding sections. Let us first briefly recall the theorem and
then go on with illustrations.

Take a probability space with a filtration $(\Omega,{\mathcal
  F},{\mathcal F}_t,p)$ and let $M_t,{\mathcal F}_t$ be a nonnegative
martingale on it such that $M_0=1$. 

If $X$ is ${\mathcal F}_s$-measurable and $t\geq s$ then basic rules
of conditional expectations yield $\Expect{XM_t}=\Expect{XM_s}$ so
that one can make a consistent definition $\tExpect{X}\equiv
\Expect{XM_t}$ whenever $X$ is ${\mathcal F}_t$ measurable. Then
$\tExpect{\cdots}$ is easily seen to be a positive linear functional
with $\tExpect{1}=1$. Hence the definition $\tilde{p}_t(A)\equiv
\tExpect{{\mathbf 1}_A}$ for $A \in {\mathcal F}_t$ makes
$(\Omega,{\mathcal F}_t,\tilde{p}_t)$ a probability space.

\vspace{.2cm}

Now take for $(\Omega,{\mathcal F},p)$ a probability space carrying a
(continuous) Brownian motion $B_t$ with its filtration ${\mathcal
  F}_t$. If $M_t$ is a continuous martingale, then It\^o's
representation formula says $M_t$ satisfies the stochastic integral
equation $M_t=1+\int_0^t M_sX_sdB_s$ for some process $X_s$ (which
satisfies a number of technical conditions, in particular $X_s$ is
${\mathcal F}_s$-measurable for each $s$). In fact one can write
$M_t=\exp \big[\int_0^t X_s\, dB_s-\frac{1}{2}\int_0^t X_s^2\,
ds\big]$ as shown by an application of It\^o's formula.
 
Girsanov's theorem states that the process $V_t\equiv B_t-\int_0^t
X_sds$ is a Brownian motion on $[0,T]$ for $(\Omega,{\mathcal
  F}_T,\tilde{p}_T)$ for each $T >0$ (see for instance \cite{Oksendal}
for a readable mathematical introduction, \cite{Karatzas} for a more
exhaustive mathematical reference or \cite{Bauer Bernard PR} for a
heuristic proof).

\vspace{.2cm}

A simple special case is $M_t\equiv e^{xB_t-tx^2/2}$, which is
a martingale on the Brownian motion space satisfying the conditions
above for a constant  $X_s=x$. Girsanov's theorem yields that under 
$\tilde{p}_T$, $V_t=B_t-xt$ is a Brownian motion on $[0,T]$, so that under
$\tilde{p}_T$ the original process $B_t$ is now a Brownian motion with a
constant drift. This makes clear that the restriction to finite $T$ 
in Girsanov's theorem is crucial: if we could take $T \rightarrow
\infty$, we would have
$B_t/t\rightarrow x$ for large $t$  with probability $1$ for
$\tilde{p}$ while this event has
probability $0$ for $p$.  

\subsection{Comparison between different kinds of SLEs}

Our computations in subsection \ref{subsec:change of domain} all dealt
with the following situation : for a certain variant of SLE on a
certain domain and with a certain normalization of the Loewner
uniformizing map the driving function satisfied a known stochastic
differential equation (in fact it was a Brownian motion), and we aimed
at the stochastic differential equation for another variant.

\vspace{.2cm}

Physical arguments allow to get this from Girsanov's theorem.  SLE
curves can be seen in general as interfaces in conformally invariant
field theories which are continuum limits of statistical mechanics
systems where an interface can already be identified in the discrete
setting due to boundary conditions.  Statistical mechanics relies on the
computation of partition functions, and in the case at hand, the
partition function when an initial segment of the interface is fixed
is (up to normalization) the probability that the interface starts
with this initial segment.  Changing the boundary conditions changes
the partition function i.e. the distribution of the interface. The
ratio of two such partition functions for the same initial segment of
interface then appears as a discrete version of a Radon-Nykodim
derivative, and indeed a slight extension of the arguments in
\cite{Bauer Bernard Houdayer} yields the fact the ratio of partition
functions is a martingale when the initial segment gets larger
\footnote{This extension however points to the importance that the
  variant of interface involved in the numerator is absolutely
  continuous with respect to the interface involved in the
  denominator.}.

A discrete partition function in an arbitrary geometry is hard to
compute, to say the least. However the conformally invariant continuum
limit (when it exists) is much more accessible.  Its relation to the
discrete system involves removing certain divergences, and the limit
in arbitrary geometry (pieces of the boundary are SLEs with some
change of boundary conditions at the tip) can be quite singular. But
ratios of partition functions behave smoothly exactly when one variant
of interface involved is absolutely continuous with respect to the
other, and the limiting ratio is expected to be the Radon-Nykodim
derivative for the corresponding probability measures in the
continuum.

\vspace{.2cm}

Let us illustrate this on an example and then say a few words on the
general framework.

The partition function $Z^{\text{dip}}_{\mathbb{D}}(x,a,b)$ for
dipolar SLE in domain $\mathbb{D}$ from a boundary point $x$ to a
boundary interval $[a,b]$ (not containing $x$) is given by a three
point correlation function
$\statav{\phi_{h_{0,1/2}}(a)\phi_{h_{1,2}}(x)\phi_{h_{0,1/2}}(b)}_{\mathbb{D}}$
where $\phi_h$ is a customary notation for a primary conformal
(boundary) field of weight $h$, and $h_{r,s}\equiv \frac{(\kappa r
  -4s)^2-(\kappa -4)^2} {16\kappa}$. So
$h_{0,1/2}=\frac{(\kappa-2)(6-\kappa)}{16\kappa}$ and
$h_{1,2}=\frac{(6-\kappa)}{2\kappa}$.

Such three point functions are completely fixed up to
normalization by conformal invariance because all quadruples
$(\mathbb{D},x,a,b)$ are conformally equivalent. If $\mathbb{D}$ is
the upper half plane $\mathbb{H}$ (then $a,b,x$ are real numbers),
\[\statav{\phi_{h}(a)\phi_{h'}(x)\phi_{h''}(b)}_{\mathbb{H}}\propto
|b-a|^{h'-h-h''}|x-a|^{h''-h-h'}
|x-b|^{h-h''-h'}.\]
More generally, if the boundary of $\mathbb{D}$ is smooth at $a,b,x$ and
$x$ and $g$ maps $\mathbb{D}$ to $\mathbb{H}$,
\[\statav{\phi_{h}(a)\phi_{h'}(x)\phi_{h''}(b)}_{\mathbb{D}}=
\statav{\phi_{h}(g(a))\phi_{h'}(g(x))\phi_{h''}(g(b))}_{\mathbb{H}}
|g'(a)|^{h}|g'(x)|^{h'}|g'(b)|^{h''},\]
emphasizing the fact that partition functions are sections of tensor
density bundles.

In the case at hand,
\[Z^{\text{dip}}_{\mathbb{H}}(x,a,b)\propto
|b-a|^{(\kappa-6)^2/(8\kappa)}|x-a|^{(\kappa-6)/(2\kappa)}
|x-b|^{(\kappa-6)/(2\kappa)}\]
and
\[Z^{\text{dip}}_{\mathbb{D}}(x,a,b)=Z^{\text{dip}}_{\mathbb{H}}
(g(x),g(a),g(b)) |g'(a)|^{\frac{(\kappa-2)(6-\kappa)}{16\kappa}}
|g'(x)|^{\frac{(6-\kappa)}{2\kappa}}
|g'(b)|^{\frac{(\kappa-2)(6-\kappa)}{16\kappa}}.\]

Chordal SLE in $\mathbb{D}$ from $x$ to $y$ has partition function
$Z^{\text{chord}}_{\mathbb{D}}(x,y)=\statav{\phi_{h_{1,2}}(x)
  \phi_{h_{1,2}}(y)}_{\mathbb{D}}$ which is even simpler. Taking
$\mathbb{D}$ to be the upper half plane $\mathbb{H}$,
\[Z^{\text{chord}}_{\mathbb{H}}(x,y)=
|x-y|^{-2h_{1,2}}=|x-y|^{(\kappa-6)/(\kappa)},\] and in general
\[Z^{\text{chord}}_{\mathbb{D}}(x,y)=
Z^{\text{chord}}_{\mathbb{H}}(g(x),g(y))
|g'(x)|^{\frac{(6-\kappa)}{2\kappa}}
|g'(y)|^{\frac{(6-\kappa)}{2\kappa}}.\] Note that if $y \rightarrow
\infty$, one has to change local coordinate to express the section,
and $Z^{\text{chord}}_{\mathbb{H}}(x,\infty)=1$.

\vspace{.2cm}
  
In the identity
\[
\frac{Z^{\text{chord}}_{\mathbb{D}}(x,y)|g'(y)|^{\frac{(\kappa-6)}{2\kappa}}}
{Z^{\text{chord}}_{\mathbb{H}}(g(x),g(y))}=
\frac{Z^{\text{dip}}_{\mathbb{D}}(x,a,b)
  |g'(a)|^{\frac{(\kappa-2)(\kappa-6)}{16\kappa}}
  |g'(b)|^{\frac{(\kappa-2)(\kappa-6)}{16\kappa}}}{Z^{\text{dip}}_{\mathbb{H}}
  (g(x),g(a),g(b))},\] the derivative of $g$ at $x$ disappears, and
the formula makes sense even if $\mathbb{D}$ is not smooth at $x$.

Suppose now that $\mathbb{D}$ is the domain obtained after the
interface has grown for some time (and $x$ is at the tip), capacity is
used to measure time, and the uniformizing map is taken in the
hydrodynamical normalization i.e. $\mathbb{D}=\mathbb{H}_t\equiv
\mathbb{H}\backslash \gamma_{]0,t]}$ is uniformized by $g_t$
satisfying $dg_t(z)=2dt/(g_t(z)-V_t)$ where $V_t\equiv
g_t(\gamma_t)$. Let $A_t\equiv g_t(a)$, $B_t\equiv g_t(b)$. Take
$y=\infty$ and observe that $g_t(\infty)=\infty$ and $g'_t(\infty)=1$.
Then the above ratio becomes
\[ Z^{\text{chord}}_{\mathbb{H}_t}(\gamma_t,\infty)=
Z^{\text{dip}}_{\mathbb{H}_t}(\gamma_t,a,b)
\frac{|g_t'(a)|^{\frac{(\kappa-2)(\kappa-6)}{16\kappa}}
  |g'_t(b)|^{\frac{(\kappa-2)(\kappa-6)}{16\kappa}}}
{Z^{\text{dip}}_{\mathbb{H}} (V_t,A_t,B_t)}\]

\vspace{.2cm}
 
Two strategies are now available. The first one looks at the ratio
\[
\frac{Z^{\text{dip}}_{\mathbb{H}_t}(\gamma_t,a,b)}
{Z^{\text{chord}}_{\mathbb{H}_t}(\gamma_t,\infty)}\equiv
\frac{Z^{\text{dip}}_{\mathbb{H}}
  (V_t,A_t,B_t)}{|g_t'(a)|^{\frac{(\kappa-2)(\kappa-6)}{16\kappa}}
  |g'_t(b)|^{\frac{(\kappa-2)(\kappa-6)}{16\kappa}}}.\] A general
argument from conformal field theory guaranties that this ration is a
martingale of chordal SLE, call it $M_t$ (of course this can also be
checked by a straightforward computation). The naive continuum limit
argument given above suggests that $M_t$ is exactly the Radon-Nykodim
derivative of dipolar SLE from $x=V_0$ to $[a,b]$ with respect to
chordal SLE.  For chordal SLE, $\kappa^{-1/2}V_t$ is a Brownian
motion, and the other building blocks in the formula for $M_t$ are of
bounded variation (i.e. do not contribute to $dV_t$ terms in $dM_t$),
so that
\[dM_t=M_t\partial_x \log Z^{\text{dip}}_{\mathbb{H}}
(x=V_t,A_t,B_t)dV_t. \] Then Girsanov's theorem yields the stochastic
differential equation of dipolar SLE:
\[dV_t=\sqrt{\kappa}dW_t+\kappa \partial_x \log
Z^{\text{dip}}_{\mathbb{H}} (x=V_t,A_t,B_t) dt, \] which coincides
with the already obtained formula.  From this viewpoint, the
appearance of a logarithmic derivative in the drift is totally
natural, and comparison with the initial formula implies an amusing
relationship between the partition function
$Z^{\text{dip}}_{\mathbb{H}} (x,a,b)$ and any conformal map $\Phi(z)$
mapping the upper-half plane to the strip $a$ to $-\infty$ and $b$ to
$+\infty$ (two such maps differ by an additive constant):
$Z^{\text{dip}}_{\mathbb{H}} (x,a,b)\Phi'(x)^{h_{1,2}}$ is
$x$-independent.

\vspace{.2cm}
 
The second one is to look at the inverse ratio 
\[
\frac{Z^{\text{chord}}_{\mathbb{H}_t}(\gamma_t,\infty)}
{Z^{\text{dip}}_{\mathbb{H}_t}(\gamma_t,a,b)}.\] 
This time CFT ensures that this is a martingale of dipolar SLE.
Assuming that dipolar is absolutely continuous with respect to chordal
i.e. that $\kappa^{-1/2}V_t$ is a Brownian motion plus a drift, this
drift is fixed by the martingale property. Of course the result is in
agreement with the other approaches. The computation is more painful
because one needs to look at the second order term in Ito's
formula. However, there is one benefit: only the martingale property
is used, and not the fact that the ratio is a Radon-Nykodim
derivative. Indeed one could replace the ratio
$Z^{\text{chord}}_{\mathbb{H}_t}(\gamma_t,\infty)/
Z^{\text{dip}}_{\mathbb{H}_t}(\gamma_t,a,b)$ 
by any correlation function of dipolar SLE in $\mathbb{H}_t$ with
hitting interval $[a,b]$, which is again a tautological martingale, to
compute the drift.

This second strategy was used in \cite{Bauer Bernard Kytola} to get
the form of the stochastic differential system describing multiple
SLE's, and in \cite{Kytola} to give the conformal field theory
approach to SLE($\kappa,\rho$'s) and generalizations thereof. Indeed,
the martingale used in section \ref{sec:sw} is the simplest partition
function\footnote{In the coulomb gas formalism of conformal field
  theory, it involves no screening charges, hence appears as a simple
  product.} describing the effect of marked points (in physics
language, these could be impurities) on SLE interfaces and was
identified as such in \cite{Kytola}. From the CFT viewpoint, this
partition function leads to an SLE$_{\kappa,\rho}$ tautological
martingale by a general argument, and no explicit computation is
required.
 
\subsection{Conditioning}

Take a probability space with a filtration $(\Omega,{\mathcal
  F},{\mathcal F}_t,p)$ and an event $J \subset \Omega$ such that  
$p(J)=\Expect{\mathbf{1}_J}  \neq 0$ and let $\tilde{p},
\tExpect{\cdots}$ be the probability and expectation on
$(\Omega,\mathcal{F})$ obtained by conditioning $p$ on $J$.

If $Y$ is a random variable, $\tilde{\mathbb{E}}(Y)=
\frac{\Expect{Y\mathbf{1}_J}} {\Expect{\mathbf{1}_J}}.$ In
particular, if $Y$ is $\mathcal{F}_s$ measurable
$\Expect{Y\mathbf{1}_J}=\Expect{\Expect{Y\mathbf{1}_J|
\mathcal{F}_s}}=\Expect{Y\Expect{\mathbf{1}_J|\mathcal{F}_s}}$.
Then $M_s\equiv \frac{\Expect{\mathbf{1}_J|\mathcal{F}_s}}
{\Expect{\mathbf{1}_J}}$ is a closed bounded positive martingale
with $M_0=1$ and $\tExpect{Y}=\Expect{YM_s}.$ From the previous
subsection, it is natural so view the numerator as a partition
function $Z_t$.

If $(\Omega,{\mathcal F},p)$ is a probability space carrying a
(continuous) Brownian motion $B_t$ with its filtration ${\mathcal
  F}_t$ and  $M_s \equiv \frac{\Expect{\mathbf{1}_J|\mathcal{F}_s}}
{\Expect{\mathbf{1}_J}}$ is a continuous martingale, Girsanov's
theorem applies.

In our illustrative example, $J$ is ``the dipolar SLE process from $0$
to the upper boundary in the strip hits in the interval
$[i\pi+a,i\pi+b]$''. We have computed the corresponding martingale:
$M_t=Z(V_t,A_t,B_t)/Z(0,a,b)$ where 
\[
Z(V,A,B)\equiv \int_{A-V}^{B-V} \, dx \cosh ^{-4/\kappa} x/2.
\]
The martingale $M_t$ is obviously continuous and Girsanov's theorem
avoids the previous elementary but clumsy discussion to yield the
drift in one stroke, and then the Markov property follows immediately. \\
Again, the processes $A_t$ and $B_t$ are differentiable with respect
to $t$ so that in the equation $M_t=1+\int_0^t M_sX_sdV_s$ only the
variation of $Z(V_s,A_s,B_s)$ with respect to $V_s$ contributes to $X_s$
and we find $X_s =\partial_V \log Z(V_s,A_s,B_s)$: the drift is again
the variation of the appropriate free energy with respect to the
starting point of the curve.

\section{Conclusions}

In this paper we have shown that if $\kappa=2$, conditioning dipolar
SLE to hit a subinterval yields dipolar SLE$_2$ towards that
subinterval. As a limiting case, conditioning dipolar SLE$_2$
to hit at a point yields chordal SLE$_2$. We have also shown that
these properties characterize the value $\kappa=2$ by computing
explicitly the drift term that arises when dipolar SLE$_\kappa$ for
general $\kappa$ is conditioned to hit a subinterval. This drift term
corresponds to some SLE$_{\kappa,\rho}$. We have shown that this drift
can be computed by standard martingale techniques or by conformal
field theory inspired methods. 

This can be seen as a characterization of $\kappa=2$ analogous
to the characterization of $\kappa=8/3$ by the restriction property or
$\kappa=6$ by locality. 

The relationship with lattice models leads to puzzling facts however.
First of all, there is no obvious interpretation of the dipolar
geometry for all lattice models. Loop erased random walks, the Ising
model for spin clusters and percolation are among the few favorable
cases. For loop erased random walks, the fact that conditioning of
dipolar to hit a subinterval is dipolar to the subinterval is true
almost by definition. For the Ising model and percolation, the
boundary condition on the hexagonal lattice are $+$ and $-$ on each
side of the starting point and \textit{free} on the boundary interval.
It is easy to understand then that conditioning to hit a subinterval
is a non trivial operation. But the case of self avoiding walks is
irritating.  The obvious candidate for the dipolar geometry would be
to consider all the simple walks joining a boundary point to a
boundary interval and weight each step by the critical fugacity
$\mu_c$. This prescription has the property that conditioning to end
at a given point yields the chordal case. So we conclude that this
naive definition cannot converge to dipolar SLE$_{8/3}$, most likely
because it corresponds to non-conformally invariant boundary
conditions.

\vspace{1cm}

\emph{Aknowledgements:} This research began during a visit to the Kavli
Institute for Theoretical Physics in September, 2006. The research of
TK was supported in part by the National Science Foundation under
grants DMS-0501168 and DMS-0758649.  The research of MB and DB was
supported in part by ANR-06-BLAN-0058.

\end{document}